\documentclass[preprint,12pt,review]{elsarticle}

\usepackage{amssymb}
\usepackage{amsmath}
\newcommand{\bi}{\bibitem}

\journal{Physics Letters B}

\begin{document}

\begin{frontmatter}



\title{The Glashow resonance in neutrino--photon scattering}


\author{I. Alikhanov\corref{cor1}}

\cortext[cor1]{{\it Email address: {\tt ialspbu@gmail.com}}}

\address{Institute for Nuclear Research of the Russian Academy of Sciences,
60-th October Anniversary pr. 7a, Moscow 117312, Russia}

\begin{abstract}
Reactions ${\nu_l}\gamma\rightarrow W^+l^-\,(l=e,\mu,\tau)$ near the threshold $\sqrt{s}=m_W+m_l$ are analyzed.
Two independent calculations of the corresponding cross sections (straightforward calculations using the Standard Electroweak Lagrangian and calculations in the framework of the parton model) are compared. It is shown that the Standard Electroweak Theory strongly suggests that these reactions proceed via the Glashow resonances. Accordingly, a hypothesis that the on-shell $W$ bosons in the reactions ${\nu_l}\gamma\rightarrow$$W^+l^-$ are the Glashow resonances is put forward. A role of these reactions for testing T~symmetry is discussed. A model with T-violating Glashow resonances for description of the distribution of the TeV-PeV neutrino events recently observed by the IceCube Collaboration is presented.
\end{abstract}


\end{frontmatter}

\section{Introduction}
In the past few decades neutrino--photon  reactions as well as their implications for astrophysics and cosmology have attracted some interest and a definite progress  has been reached in this field~\cite{chiu1960,cung1975,dicus1997,harris,Shaisultanov, abbas,Ioannisian,Aghababaie,Goyal,Chistyakov,jetp,Haghighat,newest,mine11,mine22}. For example, it has been realized that the inelastic process $\nu\gamma\rightarrow\nu\gamma\gamma$ significantly dominates over elastic scattering $\nu\gamma\rightarrow\nu\gamma$ \cite{dicus1993,dicus1999,abada1,abada2}. In its turn, when the energy threshold of the electron--positron pair production is crossed, the reaction $\nu\gamma\rightarrow\nu e^+e^-$ becomes the dominant one~\cite{masso}. 

Though neutrinos are generally considered to be weakly interacting particles, it has been shown that neutrino--photon interactions should not be confined only to discussions of loop effects in scattering, or generating neutrino magnetic moments~\cite{seckel}. In some cases $\nu\gamma$ reactions at tree level are competitive with the standard charged or neutral current neutrino scattering, and even may be dominant. An intuitive view of how a neutrino interacts with the photon is provided by the parton model~\cite{mine1,mine2}. 

With the completion of the IceCube kilometer-scale neutrino detector located at the South Pole~\cite{icecube_rev}, the idea of observing cosmic ultra-high energy (UHE) electron antineutrinos through the resonant $s$-channel reaction $\bar\nu_ee^-\rightarrow W^-$~\cite{Glashow_res,berezinsky} (the so-called Glashow resonance) is again in the focus of attention of physicists~\cite{Glashow_res_cube1, Glashow_res_cube2, Glashow_res_cube3, Glashow_res_cube4, Glashow_res_cube5,Glashow_res_cube6}. Moreover, there has already been a proposal to interpret the PeV cascade events ($\approx1.04~\text{PeV}$, $\approx1.14~\text{PeV}$, $\approx2.00~\text{PeV}$) recently reported by the IceCube experiment~\cite{icecube_2events,icecube_28,icecube_new} in terms of the Glashow resonance~\cite{Glashow_res_cube7,Glashow_res_cube8}. However, the antineutrino energy in the laboratory reference frame required to excite this resonance is  $E_{\bar\nu}\approx m_W^2/(2m_e)=6.3$ PeV (1~PeV=$10^{15}$~eV), so that the gaps in energy between the observed events and the expected resonance position are  of the order of a few PeV. It should be noticed that according to~\cite{icecube_28}, the IceCube event reconstructed energy is not due to the resonance at 6.3 PeV at 68\% C.L..

Usually in the analysis of UHE neutrino interactions, under the Glashow resonance the following reaction at $\sqrt{s}=m_W$ is implied:

\begin{eqnarray}
\bar\nu_ee^-\rightarrow W^-,
\label{introd_1}
\end{eqnarray}

though it would also be fair to refer to the remainder five similar processes predicted by the Standard Electroweak Theory, 

\begin{eqnarray}
\nu_ee^+\rightarrow W^+,\nonumber\\ \nu_ll^+\rightarrow W^+,\\
\bar\nu_ll^-\rightarrow W^-, \nonumber
\label{introd_2}
\end{eqnarray}
as to the Glashow resonances ($l=\mu,\tau$). We do so in the subsequent discussion and call any of the reactions~(1)--(2) the Glashow resonance. 

The reason for highlighting~(1) and ignoring~(2) in the literature is simply that electrons as targets are explicitly present in matter while positrons, muons and tau leptons are not. Nevertheless, we would like to remind us that one can attribute an equivalent lepton spectrum to the photon as well as to charged particles~\cite{zerwas_equiv}. Neutrinos may excite the Glashow Resonances on such equivalent leptons generated by atomic nuclei~\cite{mine1}, so that the corresponding probabilities should be studied in detail. We also emphasize that so far none of the Glashow resonances has been revealed and their experimental observation would undoubtedly be a crucial test of the Standard  Electroweak Theory.

In the present paper we analyze the reactions 

\begin{equation}
{\nu_l}\gamma\rightarrow W^+l^-,\,(l=e,\mu,\tau)
\label{introd_3}
\end{equation}
near the threshold $\sqrt{s}=m_W+m_l$~\cite{seckel}. (Our conclusions are exactly the same for the CP conjugate reactions ${\bar\nu_l}\gamma\rightarrow$$W^-l^+$  since the equivalent lepton spectrum of the photon is CP-symmetric, but for the sake of definiteness we restrict attention to (3)). 

We compare two independent calculations of the corresponding cross sections: 1) direct calculations using the Standard Electroweak Lagrangian~\cite{seckel}; 2) calculations in the framework of the equivalent particle approximation. We show that the Standard Electroweak Theory strongly suggests that the reactions (3) proceed via the Glashow resonances. Accordingly, we put forward a hypothesis that 
{\it{the on-shell $W$ bosons in the reactions ${\nu_l}\gamma\rightarrow$$W^+l^-$ are the Glashow resonances}}.

If the hypothesis is true, then the mentioned reactions provide an opportunity to observe the Glashow resonances for all neutrino flavors at laboratory energies far below 6.3~PeV. 
For example, we have found that in the reactions ${\nu_l}^{16}\text{O}\rightarrow$$^{16}\text{O} W^+l^-$, relevant for the IceCube experiment, the Glashow resonances can appear  already at neutrino energies about 20~TeV. 

A role of these reactions for testing T~symmetry at the IceCube Neutrino Observatory is discussed. We show that a model of T-violating Glashow resonance production by neutrinos interacting with the equivalent photons of the $^{16}$O nuclei is able to describe the TeV-PeV neutrino events recently observed by the  IceCube Collaboration~\cite{icecube_28}.

\section{Initial state lepton-strahlung mechanism for ${\nu_l}\gamma\rightarrow$$W^+l^-$}
\begin{figure}
\includegraphics[width=3.7in]{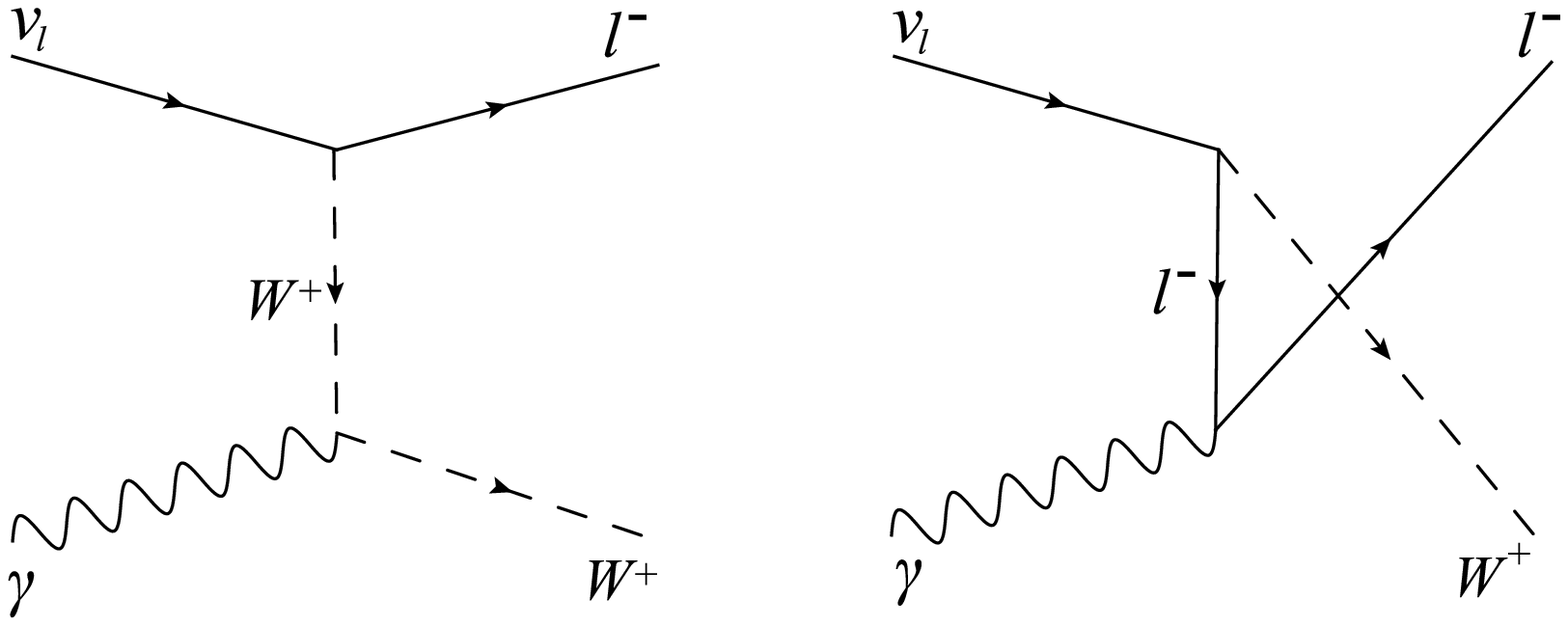}
\caption{}
\label{sm_result}
\end{figure}

\begin{figure}
\includegraphics[width=3.7in]{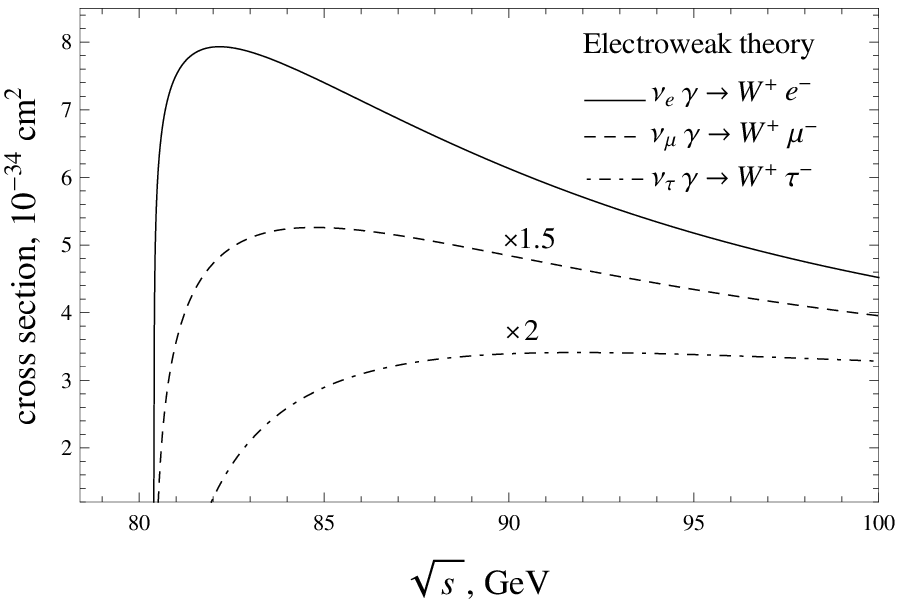}
\caption{}
\label{cross_sm}
\end{figure}

The cross sections of the reactions~(\ref{introd_3}) can be straightforwardly calculated using the Standard Electroweak Lagrangian~\cite{seckel}. The two diagrams that contribute to the amplitude at leading order are depicted in Fig.~\ref{sm_result}. The result reads

\begin{equation}
\sigma_{l}=\sqrt{2}\alpha G_F\left[2(1 - \tau)(1 + 2\tau^2 +\tau^2\log{\tau}) 
   +\tau(1 - 2\tau + 2\tau^2)\log\left(\frac{m_W^2}{m_l^2}\frac{(1-\tau)^2}{\tau}\right)\right],\label{tree}
\end{equation}

where $\tau=m_W^2/s$ and $s = (p_{\nu}+p_\gamma)^2$, $G_F$ is the Fermi
constant, and $\alpha$ is the fine structure constant. Figure~\ref{cross_sm} shows the cross sections for
the three different neutrino flavors.

\begin{figure}
\includegraphics[width=3.7in]{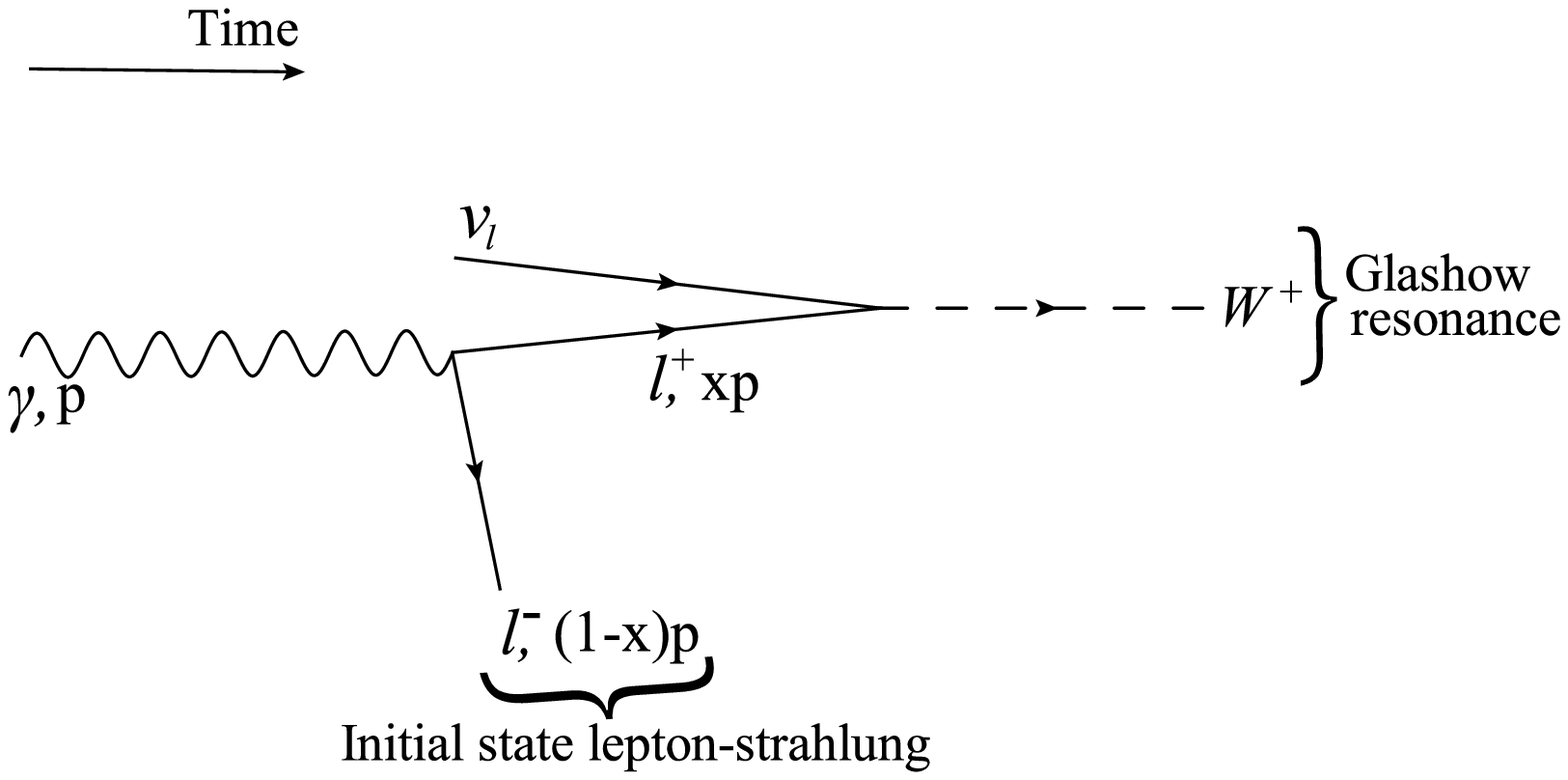}
\caption{}
\label{s_chan}
\end{figure}

\begin{figure}
\includegraphics[width=3.7in]{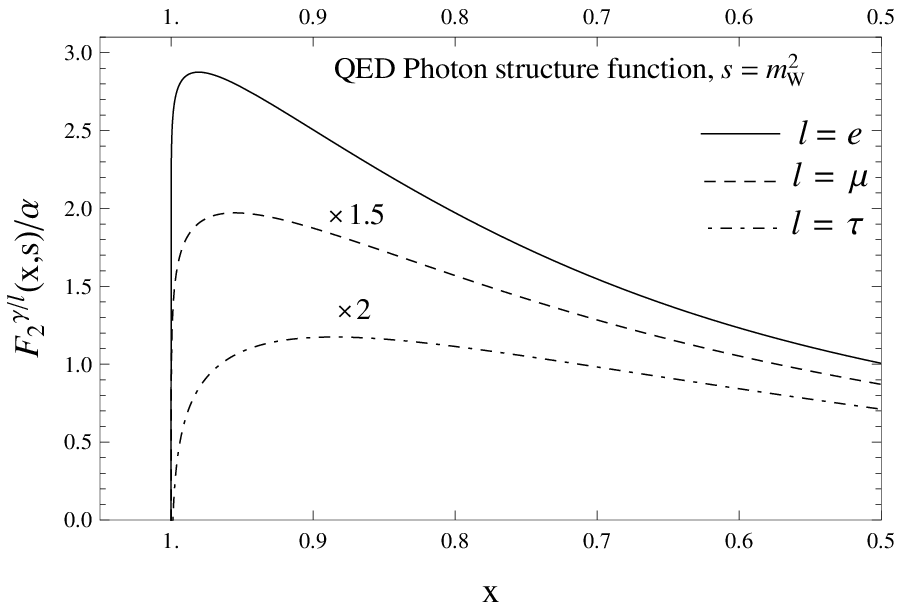}
\caption{}
\label{struc_fun}
\end{figure}

One may notice the sharp rise of the cross sections at $\sqrt{s}\approx m_W+m_l$ (especially for $\nu_e$) and the subsequent slow falling with energy. This is typical for processes in which the so-called initial state radiation takes place. It is well known that emission of real or virtual photons from the initial colliding electrons essentially modify the shapes of the narrow resonance curves~\cite{lipatov}: the curves become wider, a suppression of the resonance maximum is observed and the main distinctive feature -- {\it the radiation tail} -- appears to the right of the resonance pole. The matter is that even if the collision energy $\sqrt{s}$ exceeds the mass of the resonance $m_R$, the radiated photon carries away the energy excess $E_\gamma=\sqrt{s}-m_R$ before $e^+e^-$ annihilation and thus turns back the $e^+e^-$ pair to the resonance energy. 
\newline Analogously, it is tempting to identify the shapes of the cross sections in Fig.~\ref{cross_sm} with the radiation tails arising due to initial state emission of charged leptons from the photon (initial state lepton-strahlung).  In order to do this, we have to assume the following mechanism for the reactions~(\ref{introd_3}) schematically illustrated in Fig~\ref{s_chan}:  the initial photon splits into a $l^+l^-$ pair and subsequently the positively charged lepton from this pair annihilates with the ingoing neutrino into $W^+$ (the Glashow resonance), while the energy excess $\sqrt{s}-m_W$ is carried away by the outgoing $l^-$. 
\newline In addition to the peculiarities of the behavior of the cross sections near and above the threshold, there is also another argument strongly suggesting the initial state lepton-strahlung mechanism for ${\nu_l}\gamma\rightarrow W^+l^-$. Let us plot the QED structure functions of the photon, $F^{\gamma/l}_{2}(x,s)$, in a graph with flipped abscissa (recall that  $F^{\gamma/l}_{2}(x,s)/x$ gives the probability density of finding a charged lepton in the photon with fraction $x$ of the parent photon's momentum). When looking at such a graph shown in Fig.~\ref{struc_fun}, one immediately recognizes the similarity to the shapes of cross sections from Fig~\ref{s_chan}. It should be emphasized that the structure functions are obtained independently for deep inelastic charged lepton--photon scattering~\cite{nisius}. An explanation for this similarity is that the relatively narrow Glashow resonances project out the structure functions of the photon $F^{\gamma/l}_{2}(x,s)$ in the cross sections $\sigma_l$~\cite{zerwas_equiv}. Formally this is well seen by exploiting the parton model approach which tells us that $\sigma_l$ can be written as

\begin{equation}
\sigma_{l}=\int \frac{dx}{x}F^{\gamma/l}_{2}(x,s)\sigma_{\nu l\rightarrow W}(xs),
\label{eq_main_convol}
\end{equation}

where $\sigma_{\nu l\rightarrow W}(xs)$ are the cross sections of the subprocesses $\nu_ll^+\rightarrow W^+$, the integration is performed over the kinematically allowed values of $x$.
In the narrow width approximation $\sigma_{\nu l\rightarrow W}(xs)$ can be replaced by a Dirac $\delta$~function, so that $\sigma_{\nu l\rightarrow W}(xs)=2\sqrt{2}\pi G_F \tau\delta(x-\tau)$. Substituting the latter into~(\ref{eq_main_convol}) yields

\begin{equation}
\sigma_{l}=2\sqrt{2}\pi G_F F^{\gamma/l}_{2}(\tau,s).
\label{eq_narrow2}
\end{equation}

Thus, one can conclude from~(\ref{eq_narrow2}) that our mechanism accounts for the similarity between $\sigma_l$ and $F^{\gamma/l}_{2}(x,s)$: they turn out to be proportional to each other.
Analytically, $F^{\gamma/l}_{2}(x,s)$ is parametrized as~\cite{nisius} 

\begin{equation}
F^{\gamma/l}_{2}(x,s)=\frac{\alpha}{2\pi}x\Biggl[8x(1-x)-1+[x^2+(1-x)^2]\log\left(\frac{s(1-x)}{m_l^2}\right)\Biggr]
\label{eq_narrow}
\end{equation}

\begin{figure}
\includegraphics[width=3.7in]{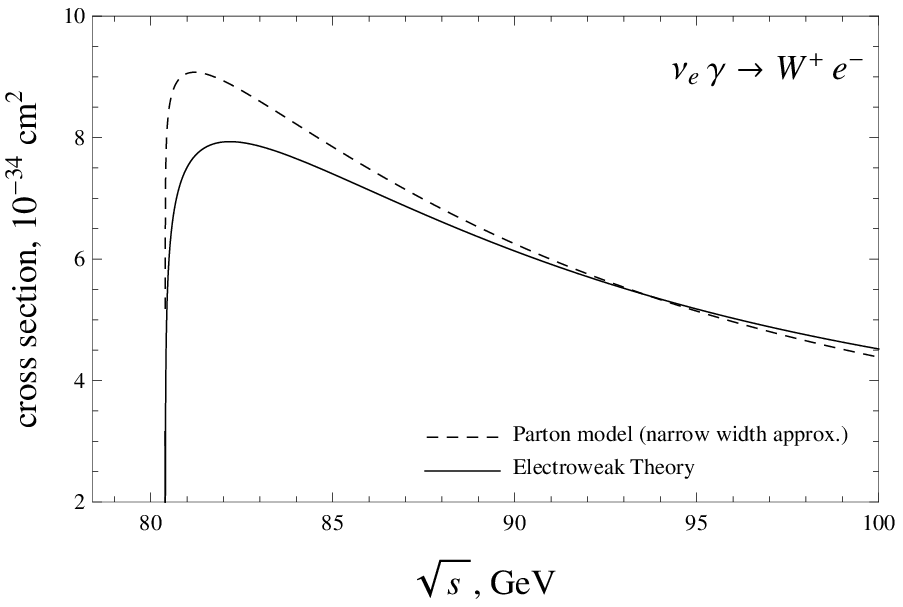}
\caption{}
\label{narrow}
\end{figure}

As an example, Fig.~\ref{narrow} shows the cross section for the reaction ${\nu_e}\gamma\rightarrow$$W^+e^-$ given by~(\ref{eq_narrow2}) with~(\ref{eq_narrow}) and that taken from~\cite{seckel}. The fact that the independently obtained structure function reproduces the straightforward Standard Electroweak Theory calculations within the error $<20\%$ inspires to go into further details.

Let us utilize the cross sections for the subprocesses $\nu_ll^+\rightarrow W^+$ in the Breit--Wigner form making the description of the resonances more physically realistic than the narrow width approximation. In this case, one has

\begin{equation}
\sigma_{l}=24\pi\mathrm{\Gamma_{W\rightarrow\nu{\it l}}}\mathrm{\Gamma}\int_{x_{\text{min}}}^{x_{\text{max}}}\frac{dx}{x}\frac{F^{\gamma/l}_{2}(x,s)}{(xs-m_W^2)^2+m_W^2\mathrm{\Gamma}^2},\label{eq_convol}
\end{equation}

where $\mathrm{\Gamma_{W\rightarrow\nu{\it l}}}$ is the width of the decay $W^+\rightarrow\nu_l l^+$, $\mathrm{\Gamma}$~is the total
decay width of $W^+$. $x_{\text{min}}=m_l^2/s$, $x_{\text{max}}=(1-m_l/\sqrt{s})^2$.

\begin{figure}
\includegraphics[width=3.7in]{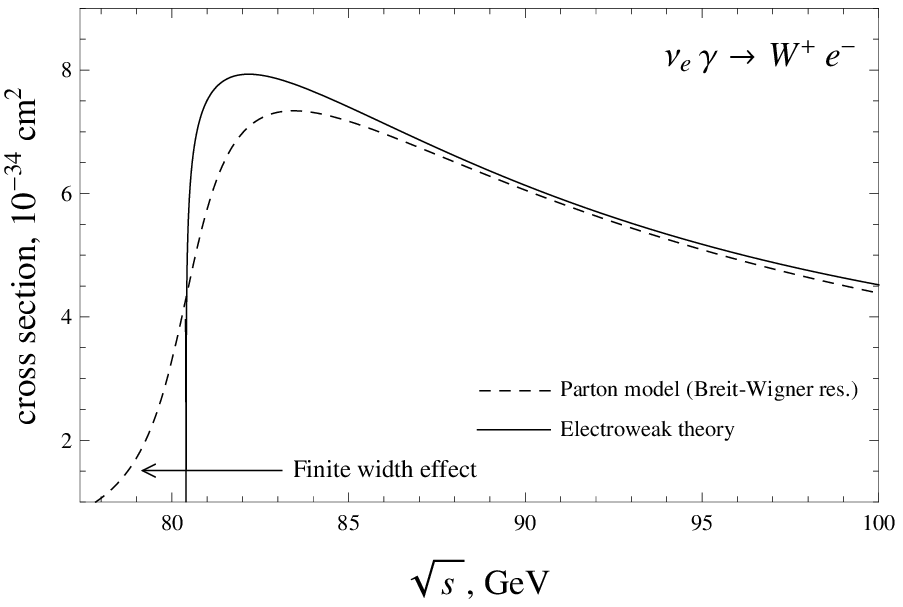}
\caption{}
\label{pm_el}
\end{figure}

\begin{figure}
\includegraphics[width=3.7in]{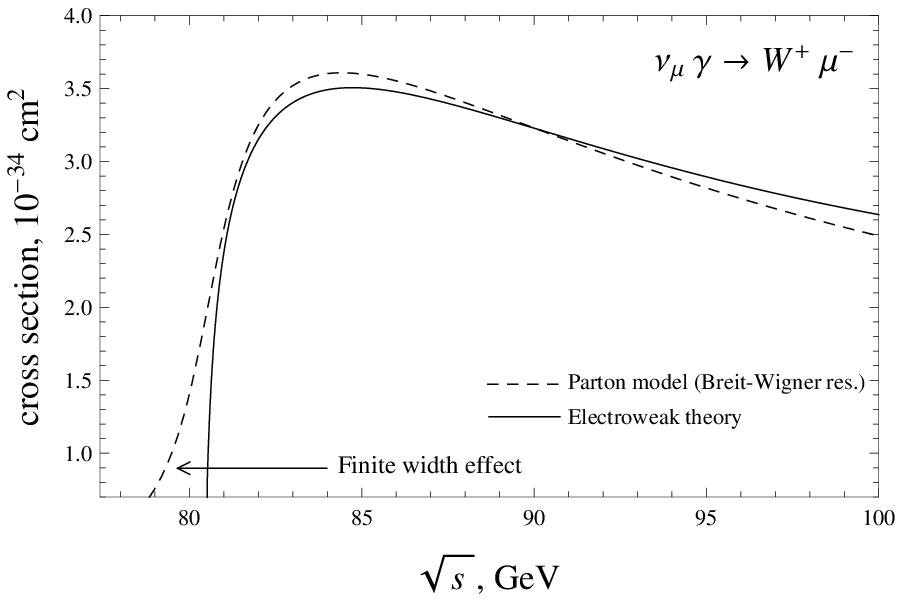}
\caption{}
\label{pm_mu}
\end{figure}

\begin{figure}
\includegraphics[width=3.7in]{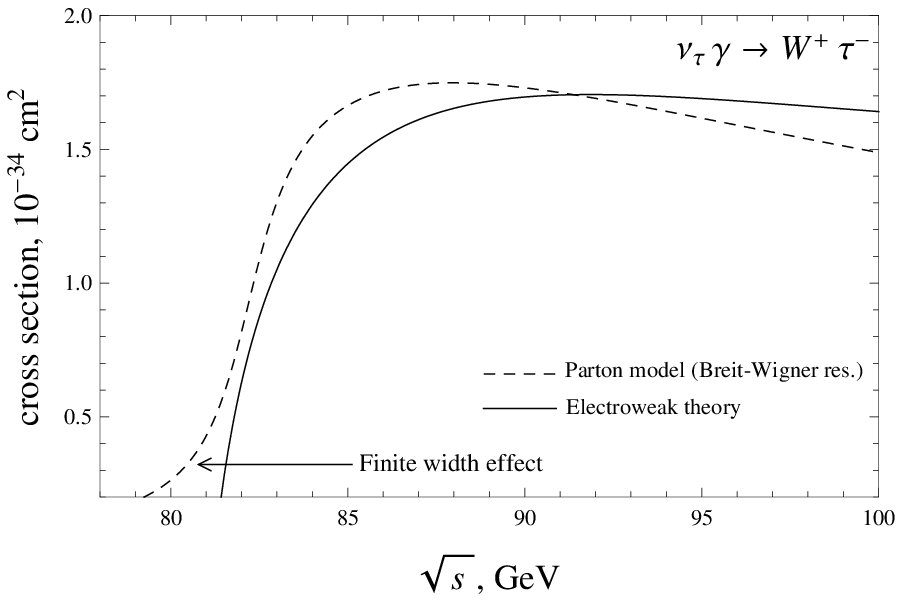}
\caption{}
\label{pm_tau}
\end{figure}

Substituting~(\ref{eq_narrow}) into~(\ref{eq_convol}), we have performed calculations and display the results in comparison with the direct Standard Electroweak Theory predictions in Figs.~\ref{pm_el}, \ref{pm_mu} and \ref{pm_tau}. One can see that our model is again in a very good quantitative agreement with the straightforward standard calculations~\cite{seckel}. The shifts of the reaction thresholds to energies slightly lower than $\sqrt{s}=m_W+m_l$ are obviously due to the finite width effect~\cite{finite_w}: the W propagator adopted in~\cite{seckel} has the structure $\sim1/(q^2-m_W^2)$, while we have used it in the general form $\sim1/(q^2-m_W^2+im_W\Gamma)$. 
Everywhere in the calculations we have taken $\alpha=1/128$, $G_F=1.16\times10^{-5}$ $\text{GeV}^{-2}$,   $m_e=5\times10^{-4}$ GeV, $m_{\mu}=0.105$ GeV, $m_{\tau}=1.77$ GeV, $m_W=80.4$ GeV, $\Gamma=2.14$ GeV, $\Gamma_{\nu{\it l}}=0.23$ GeV.

\section{The Glashow resonance in neutrino--nucleus scattering}
The Glashow resonances can be produced in neutrino--nucleus scattering $\nu_lN\rightarrow Nl^-W^+$.  To find the corresponding cross sections one has to convolute~(\ref{eq_convol}) with the equivalent photon spectrum of the nucleus:

\begin{equation}
\sigma_{Nl}=\int_{y_{0}}^1 dyf^{N/\gamma}(y)\sigma_l(ys),
\label{nucleus}
\end{equation}
where $y_0=(m_W+m_l)^2/s$, $f^{N/\gamma}(y)$ is the equivalent photon spectrum. The general expression for $f^{N/\gamma}(y)$ can be written as~\cite{serbo}

\begin{equation}
f^{N/\gamma}(y)=\frac{\alpha Z^2}{2\pi}\frac{2-2y+y^2}{y}\int_{Q^2_{\text{min}}}^{\infty}dQ^2\frac{Q^2-Q^2_{\text{min}}}{Q^4}\left|F(Q^2)\right|^2,
\label{photon_flux}
\end{equation}
where $Q^2$ is the momentum transfer to the nucleus, $Z$ is the charge of the nucleus, $F(Q^2)$ is the electromagnetic nuclear formfactor normalized to $F(0)=1$, $Q_{\text{min}}=(yM_N)^2/(1-y)$ with $M_N$ being the mass of the nucleus.  

\begin{figure}
\includegraphics[width=3.7in]{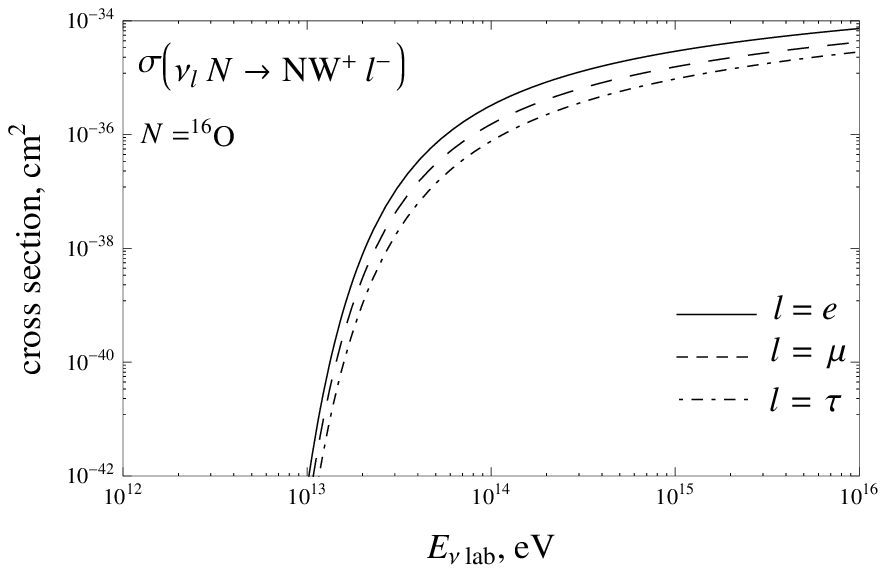}
\caption{}
\label{nucl_cross}
\end{figure}

Let us consider the nucleus of oxygen $^{16}\text{O}$ as the target because ultra high energy neutrinos can be detected  in large volumes of water or ice, for example, the IceCube kilometer-scale detector~\cite{icecube_rev},  the ANTARES undersea neutrino telescope~\cite{antares} as well as the next generation deep-water neutrino telescopes KM3NeT~\cite{km3net} and NT1000 on Lake Baikal~\cite{baikal}. Figure~\ref{nucl_cross} shows our calculations of the cross sections for the three neutrino flavors as functions of neutrino energy in the laboratory reference frame. The formfactor of $^{16}\text{O}$ was taken from~\cite{o16}. At the same neutrino energy, the contribution to the cross section from elastic neutrino--proton scattering will be much less than that from the coherent neutrino--nucleus interactions due to $f^{N/\gamma}\sim 1/y$, where $y\approx m_W^2/(2M_NE_{\nu})$.

\section{Violation of T symmetry?}

\begin{figure}
\includegraphics[width=3.7in]{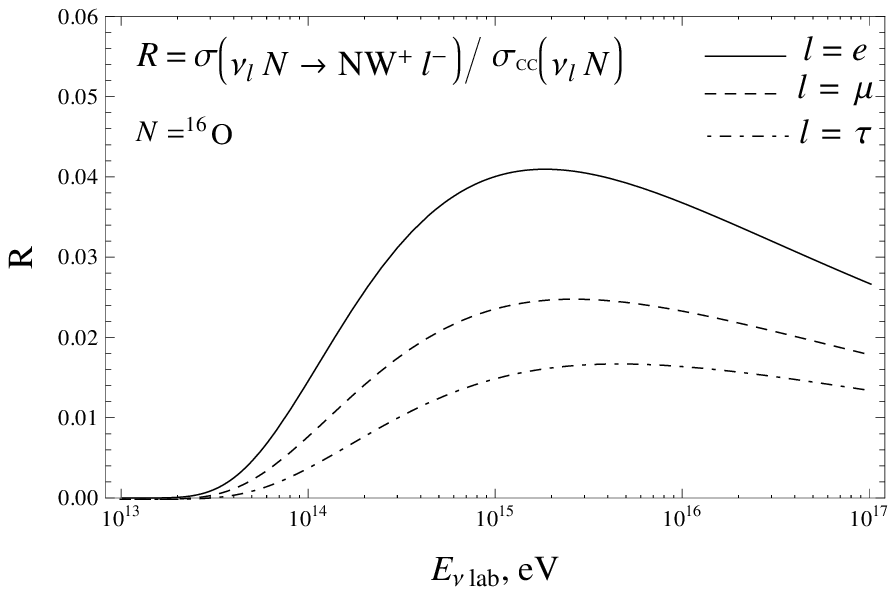}
\caption{}
\label{fig_ratio}
\end{figure}

Figure~\ref{fig_ratio} shows the
ratio of the cross-sections per nucleon for ${\nu_l}^{16}\text{O}\rightarrow$$^{16}\text{O} W^+l^-$ found in the previous section, to that for the charged current neutrino--nucleus scattering~\cite{cc_cross} as a function of neutrino energy in the laboratory reference frame.  One can see that the neutrino detection rate due to the Glashow resonance production is expected to increase by $\sim4\%$ in the energy range from $\sim20$ TeV to several PeV. This is about two times lower than the result of similar calculations performed in~\cite{seckel}, where the existence of this enhancement was pointed out for the first time. This discrepancy may be due to different treating the nuclear formfactor. It should be emphasized that the errors in calculating $\sigma_{CC}(\nu_lN)$ due to uncertainties on the parton distribution functions of the nucleon do not exceed 3\% in the neutrino energy range considered~\cite{cc_cross,pdf_error}.  One can also notice that the energy range is essentially overlapping with the energies of the TeV--PeV neutrino candidates recently announced by the IceCube Collaboration. In a three-year dataset the IceCube observed 37 events with in-detector deposited energies between $30$~TeV and 2 PeV~\cite{icecube_28,icecube_new}. 

In the view of the overlap between our calculations and the IceCube data we return to the formula~(\ref{eq_convol}) and notice that by writing~(\ref{eq_convol}) we have implicitly exploited the principle of detailed balance related to T symmetry. The matter is that we have taken 

\begin{equation}
\mathrm{\Gamma_{\nu{\it l}\rightarrow W}}=\mathrm{\Gamma_{W\rightarrow\nu{\it l}}}.
\end{equation}

Of course, the experimental value of the decay width $\mathrm{\Gamma_{W\rightarrow\nu{\it l}}}$ is in agreement with the Standard Electroweak Model~\cite{pdg}, however $\mathrm{\Gamma_{\nu{\it l}\rightarrow W}}$ was never experimentally measured. Therefore, if we are strict, the possibility that 

\begin{equation}
\mathrm{\Gamma_{\nu{\it l}\rightarrow W}}\neq\mathrm{\Gamma_{W\rightarrow\nu{\it l}}}
\end{equation}

should not be rejected at present. Then, we have to multiply~(\ref{eq_convol}) by the following correction factor:

\begin{equation}
g_l=\frac{\mathrm{\Gamma_{\nu{\it l}\rightarrow W}}}{\mathrm{\Gamma_{W\rightarrow\nu{\it l}}}},
\end{equation}

which takes into account eventual T violation in the Glashow resonances.

The expected event rate distribution at IceCube reads

\begin{equation}
\frac{dN}{dE_{\nu}}=nt\Omega\sum_{l=e,\mu,\tau}g_l\sigma_{Nl}\Phi_{\nu_l+\bar\nu_l},\label{eq_distribe}
\end{equation}

where $n$ is the number of target nuclei in the effective volume of the detector, $t$ is the time of exposure, $\Omega$ is the solid angle, $\Phi_{\nu+\bar\nu}$ is the flux of neutrinos and antineutrinos of flavor $l$. Taking $n=1.3\times10^{37}$ (the number of the $^{16}\text{O}$ nuclei in the effective volume 0.44 km$^3$), $t=365$ days, assuming the 1:1:1 neutrino flavor ratio (this is not inconsistent with the present IceCube data though there are arguments against the equal flavor composition~\cite{1404}), neglecting the upward going electron and muon neutrinos (because they are very likely absorbed in the Earth), adopting $\Phi_{\nu+\bar\nu}=\Phi_0(E_{\nu}/1\text{GeV})^{-2.3}$, $\Phi_0=6.62\times10^{-7}/(\text{GeV}\text{cm}^{2}\text{s}\,\text{sr})$~\cite{1312}, we have estimated the number of the Glashow resonance events per year at IceCube in the energy range from 15~TeV to 2~PeV:

\begin{equation}
N_{\text{t}}=\left(0.09g_{e}+0.04g_{\mu}+0.05g_{\tau}\right)\left(\frac{\Phi_0}{\frac{6.62\times10^{-7}}{\text{GeV}\text{cm}^{2}\text{s}\,\text{sr}}}\right)\left(\frac{V}{0.44~\text{km}^3}\right)\,\text{yr}^{-1}.
\label{estim}
\end{equation}

Thus, experimental investigations of the neutrino events in this energy range automatically probe T symmetry in Glashow resonance production. 

\begin{figure}
\includegraphics[width=3.7in]{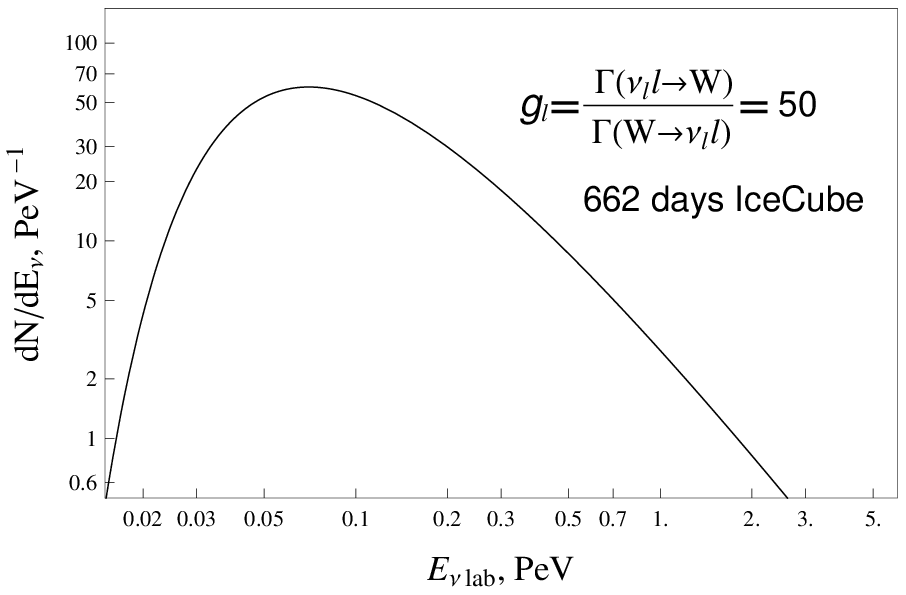}
\caption{}
\label{events}
\end{figure}

It is interesting that our approach with T-violating Glashow resonances is able to account for the TeV-PeV IceCube events~\cite{icecube_28}. In Fig.~\ref{events} we show the event rate distribution at IceCube according to~(\ref{eq_distribe}) for the time of exposure equal to 662 days and $g_{e}=g_{\mu}=g_{\tau}=50$.  This gives 18 events in total in the range from 15~TeV to 2~PeV. Since high-momentum transfer neutrino interactions with the equivalent photons is suppressed by the nuclear formfactor, almost all the incident neutrino momentum will be transferred to the W boson and either totally or by half (depending on the W boson decay mode) will be deposited in the detector. The ratio of the track events to the shower events in the detector is then obviously given by

\begin{equation}
\frac{\text{tracks}}{\text{showers}}=\frac{\mathrm{\Gamma_{W\rightarrow\nu{\it \mu}}}+0.18\Gamma_{W\rightarrow\nu{\it \tau}}}{\Gamma_{W\rightarrow\text{hadrons}}+\Gamma_{W\rightarrow\nu{\it e}}+0.82\Gamma_{W\rightarrow\nu{\it \tau}}}\approx0.15,
\end{equation}

where we take into account the subsequent decay of $\tau$ into a muon plus the corresponding (anti)neutrinos whose branching fraction is about 0.18. 

Meanwhile, the IceCube reports a two times higher value~\cite{icecube_new}:

\begin{equation}
\left.\frac{\text{tracks}}{\text{showers}}\right|_{\text{exp}}=\frac{8}{28}\approx0.28.\label{icecube_ration}
\end{equation}

Nevertheless, one should keep in mind that the numerator of~(\ref{icecube_ration}) includes the background muon tracks whose expected number is not less than 3~\cite{icecube_28}, so that the experimental ratio $\text{tracks}/\text{showers}\lesssim5/28\approx0.18$.

\section{Conclusions}
We have analyzed the reactions 
${\nu_l}\gamma\rightarrow W^+l^-,\,(l=e,\mu,\tau)$ near the threshold $\sqrt{s}=m_W+m_l$.
We compare two independent calculations of the corresponding cross sections: 1) direct calculations using the Standard Electroweak Lagrangian~\cite{seckel}; 2) calculations in the framework of the equivalent particle approximation. We show that the Standard Electroweak Theory strongly suggests that these reactions proceed via the Glashow resonances. In more detail, the analysis indicates the following mechanism for these reactions:  the initial photon splits into a $l^+l^-$ pair and subsequently the positively charged lepton from this pair annihilates with the ingoing neutrino into $W^+$ (the Glashow resonance), while the energy excess $\sqrt{s}-m_W$ is carried away by the outgoing $l^-$. It is essential that the leptons are radiated before the Glashow resonance appears. We call this mechanism "initial state lepton-strahlung".

Accordingly, we put forward a hypothesis that the on-shell $W$ bosons in the reactions ${\nu_l}\gamma\rightarrow$$W^+l^-$ are the Glashow resonances.

If the hypothesis is true, then the mentioned reactions provide an opportunity to observe the Glashow resonances for all neutrino flavors at the laboratory energies far below 6.3~PeV. 
For example, we have found that in the reactions ${\nu_l}^{16}\text{O}\rightarrow$$^{16}\text{O} W^+l^-$, relevant for the IceCube experiment, the Glashow resonances can appear  already at neutrino energies about 20~TeV. 

It turns out that the Standard Model predicts a somewhat enhancement of the Glashow resonance event rate in ice in the energy region, where the IceCube Collaboration has detected 28 neutrino candidates~ with energies from 30 TeV to 1.2 PeV\cite{icecube_28}, which is about two times higher than the signal expected from the atmospheric neutrinos. We show that experimental investigations of the neutrino events in this energy range automatically probe T symmetry in Glashow resonance production.
Relying on this observation we show that a model with T-violating Glashow resonances produced by neutrinos interacting with the equivalent photons of the $^{16}$O nuclei  is able to account for the distribution of these neutrino events at IceCube.

Our conclusions as well as numerical results are exactly the same for the CP conjugate reactions ${\bar\nu_l}\gamma\rightarrow$$W^-l^+$  since the equivalent lepton spectrum of the photon is assumed to be CP-symmetric. 

We would also like to note that there are processes described by diagrams whose structures at tree level coincide with that for ${\nu_l}\gamma\rightarrow W^+l^-$. For example, one encounters such diagrams in single scalar and vector leptoquark production in lepton--gluon scattering $l({\nu_l})+\text{g}\rightarrow LQ+q$~\cite{spira,mine_leptoq} which also lead to the cross sections with the feature resembling the radiation tail. Therefore, it is also fair to expect that they proceed, in analogy with the initial state lepton-strahlung, via an initial state quark-strahlung mechanism and the leptoquarks in these reactions are produced in $s$-channel subprocesses. Additionally, the W boson in the so-called decay of the UHE neutrino in a magnetic field $\nu_e\rightarrow W^+e^-$~\cite{kuznetsov} probably appears through the resonant annihilation subprocess $\nu_ee^+\rightarrow W^+$ as well.

\vskip 0.5cm

{\bf Acknowledgements}
\vskip 0.5cm

This work was supported in part by the Russian Foundation for Basic Research (grant 11-02-12043), by the Federal Target Program  of the Ministry of Education and Science of Russian Federation "Research and Development in Top Priority Spheres of Russian Scientific and Technological Complex for 2007-2013" (contract No. 16.518.11.7072) and by the Program for Basic Research of the Presidium of the Russian Academy of Sciences "Fundamental Properties of Matter and Astrophysics".


\newpage

{\bf Figure Captions}
\vskip 0.5 cm 

{\bf Fig. 1:} Diagrams that contribute to the amplitude for ${\nu_l}\gamma\rightarrow$$W^+l^-$ at leading order~\cite{seckel}.

\vskip 0.5 cm

{\bf Fig. 2:} Cross sections for ${\nu_l}\gamma\rightarrow$$W^+l^-$ as functions of the center-of-mass energy  $\sqrt{s}$ straightforwardly calculated in the Standard Electroweak Theory~\cite{seckel}.

\vskip 0.5 cm

{\bf Fig. 3:} A schematic illustration of the initial state lepton-strahlung mechanism of Glashow resonance production in ${\nu_l}\gamma\rightarrow$$W^+l^-$. The photon with a four-momentum $p$ splits into a $l^+l^-$ lepton pair  before the Glashow resonance emerges ($x$ is the fraction of the parent photon's momentum carried by the positively charged lepton). Even if the center-of-mass energy of the $\nu_l\gamma$ collision $\sqrt{s}$ exceeds the mass of the resonance $m_W$, the radiated $l^-$ carries away the energy excess $(1-x)s=s-m_W^2$ and turns back the $\nu_ll^+$ pair to the resonance pole  $xs=m_W^2$.



\vskip 0.5 cm

{\bf Fig. 4:} QED structure functions of the photon divided by $\alpha$ for three charged leptons~\cite{nisius}. Note that the abscissa is flipped.



\vskip 0.5 cm

{\bf Fig. 5:} Cross sections for ${\nu_e}\gamma\rightarrow$$W^+e^-$ as functions of the center-of-mass energy $\sqrt{s}$  straightforwardly calculated in the Standard Electroweak Theory (solid)~\cite{seckel} and found in the parton model with the narrow width approximation of the Glashow resonance (dashed). 

\vskip 0.5 cm

{\bf Fig. 6:} Cross sections for ${\nu_e}\gamma\rightarrow$$W^+e^-$ as functions of the center-of-mass energy $\sqrt{s}$ straightforwardly calculated in the Standard Electroweak Theory (solid)~\cite{seckel} and in the parton model with the Breit--Wigner form of the Glashow resonance (dashed).

\vskip 0.5 cm

{\bf Fig. 7:} Cross sections for ${\nu_{\mu}}\gamma\rightarrow$$W^+\mu^-$ as functions of the center-of-mass energy $\sqrt{s}$ straightforwardly calculated in the Standard Electroweak Theory (solid)~\cite{seckel} and in the parton model with the Breit--Wigner form of the Glashow resonance (dashed).



\vskip 0.5 cm

{\bf Fig. 8:} Cross sections for ${\nu_{\tau}}\gamma\rightarrow$$W^+\tau^-$ as functions of the center-of-mass energy $\sqrt{s}$ straightforwardly calculated in the Standard Electroweak Theory (solid)~\cite{seckel} and in the parton model with the Breit--Wigner form of the Glashow resonance (dashed).

\vskip 0.5 cm


\vskip 0.5 cm

{\bf Fig. 9:} Cross sections per nucleon for ${\nu_l}^{16}\text{O}\rightarrow$$^{16}\text{O} W^+l^-$ as functions of neutrino energy in the laboratory reference frame.

\vskip 0.5 cm

{\bf Fig. 10:} Ratio of the cross sections for ${\nu_l}^{16}\text{O}\rightarrow$$^{16}\text{O} W^+l^-$ to that for charged  current neutrino--nucleus scattering~\cite{cc_cross}.  The cross sections are per nucleon.

\vskip 0.5 cm

{\bf Fig. 11:} Event rate distribution $dN/dE_{\nu}$ at IceCube in the model with T-violating  Glashow resonances ($g_e=g_{\mu}=g_{\tau}=50$). The time of exposure equals to 662 days. The neutrino flux is assumed to fall with energy as $E^{-2.3}_{\nu}$.


\newpage

\end{document}